\documentstyle[preprint,eqsecnum,aps,epsfig]{revtex}
\tightenlines

\begin{document}

\draft  

\preprint{CWRU-P5-00/UCSB-HEP-00-01}

\title{Exclusion Limits on the WIMP-Nucleon Cross-Section from \\
the Cryogenic Dark Matter Search}

\author{
R.~Abusaidi$^8$,
D.S.~Akerib$^1$,
P.D.~Barnes,~Jr.$^9$, 
D.A.~Bauer$^{10}$, 
A.~Bolozdynya$^1$,
P.L.~Brink$^8$,
R.~Bunker$^{10}$,
B.~Cabrera$^8$,
D.O.~Caldwell$^{10}$, 
J.P.~Castle$^8$,
R.M.~Clarke$^8$,
P.~Colling$^8$,
M.B.~Crisler$^2$,
A.~Cummings$^9$,
A.~Da~Silva$^9$, 
A.K.~Davies$^8$,
R.~Dixon$^2$,
B.~Dougherty$^8$,
D.~Driscoll$^1$,
S.~Eichblatt$^2$,
J.~Emes$^3$,
R.J.~Gaitskell$^9$, 
S.R.~Golwala$^{9,}$\cite{byline},
D.~Hale$^{10}$,
E.E.~Haller$^3$, 
J.~Hellmig$^9$,
M.E.~Huber$^{11}$,
K.D.~Irwin$^4$, 
J.~Jochum$^9$, 
F.P.~Lipschultz$^7$,
V.~Mandic$^9$,
J.M.~Martinis$^4$,
S.W.~Nam$^4$, 
H.~Nelson$^{10}$,
B.~Neuhauser$^7$,
M.~Penn$^8$,
T.A.~Perera$^1$,
M.C.~Perillo~Isaac$^9$,
B.~Pritychenko$^9$,
R.R.~Ross$^{3,9}$, 
T.~Saab$^8$,
B.~Sadoulet$^{3,9}$,
R.W.~Schnee$^1$, 
D.N.~Seitz$^9$,
P.~Shestople$^7$,
T.~Shutt$^5$,
A.~Smith$^3$,
G.W.~Smith$^9$,
A.H.~Sonnenschein$^{10}$,
A.L.~Spadafora$^9$,
W.~Stockwell$^9$,
J.D.~Taylor$^3$,
S.~White$^9$,
S.~Yellin$^{10}$,
B.A.~Young$^6$ \\
\smallskip
(CDMS Collaboration) \\
\smallskip
}


\address{$^1$Department of Physics, Case Western Reserve University,   
		  Cleveland, OH 44106, USA   }
\address{$^2$Fermi National Accelerator Laboratory,  
                  Batavia, IL 60510, USA   }
\address{$^3$Lawrence Berkeley National Laboratory,  
                  Berkeley, CA 94720, USA   }
\address{$^4$National Institute of Standards and Technology,  
                  Boulder, CO 80303, USA   }
\address{$^5$Department of Physics, Princeton University,  
                  Princeton, NJ 08544, USA   }
\address{$^6$Department of Physics, Santa Clara University,  
	          Santa Clara, CA 95053, USA   }
\address{$^7$Department of Physics and Astronomy, 
                  San Francisco State University,  
		  San Francisco, CA 94132, USA   }
\address{$^8$Department of Physics, Stanford University,  
                  Stanford, CA 94305, USA   }
\address{$^9$Center for Particle Astrophysics, 
		  University of California, Berkeley,  
                  Berkeley, CA 94720, USA   }
\address{$^{10}$Department of Physics, 
                  University of California, Santa Barbara,   
	          Santa Barbara, CA 93106, USA   }
\address{$^{11}$Department of Physics, University of Colorado,   
		 Denver, CO 80217, USA  }

\date{\today}
\maketitle
\begin{abstract}
The Cryogenic Dark Matter Search (CDMS) employs Ge and Si detectors to
search for WIMPs via their elastic-scattering interactions with nuclei
while discriminating against interactions of background
particles.  CDMS data give limits on the spin-independent WIMP-nucleon
elastic-scattering cross-section that exclude unexplored parameter
space above 10~GeV~c$^{-2}$ WIMP mass and, at $> 84$\% CL, the entire
3$\sigma$ allowed region for the WIMP signal reported by the DAMA
experiment.
\end{abstract}


\narrowtext

Extensive evidence indicates that a large fraction of the matter in
the universe is nonluminous, nonbaryonic and ``cold'' --
nonrelativistic at the time matter began to dominate the energy
density of the universe\cite{trimble,peebles,olive}.  Weakly
Interacting Massive Particles (WIMPs) are an excellent candidate for
nonbaryonic, cold dark matter~\cite{lee,peebles}.  Minimal
supersymmetry provides a natural WIMP candidate in the form of the
lightest superpartner, with a typical mass $M \sim
100$~GeV~c$^{-2}$~\cite{jkg,ellis,gondolo,bottino}.  WIMPs are expected to
have collapsed into a roughly isothermal, spherical halo within which
the visible portion of our galaxy resides.  WIMPs scatter off nuclei
via the weak interaction, potentially allowing their direct detection
\cite{goodman,primack}.  The expected spectrum of recoil energies
(energy given to the recoiling nucleus during the interaction) is
exponential with a characteristic energy of a few to tens of keV
\cite{lewin}.  The expected event rate is model-dependent but is
generically 1~kg$^{-1}$~d$^{-1}$ or lower~\cite{primack}.

This Letter reports new exclusion limits on the spin-independent
WIMP-nucleon elastic-scattering cross-section by the Cryogenic Dark
Matter Search (CDMS).  The rate of rare WIMP-nucleon interactions is
constrained by extended exposure of detectors that
discriminate WIMP-induced nuclear recoils from electron recoils
caused by interactions of background particles
\cite{tomprl1,tomprl2}.

The ionization yield $Y$ (the ratio of ionization production to recoil
energy in a semiconductor) of a particle interaction differs greatly
for nuclear and electron recoils.  CDMS detectors measure phonon and
electron-hole pair production to determine recoil energy and
ionization yield for each event.  The data discussed here were
obtained with two types of detectors, Berkeley Large Ionization- and
Phonon-mediated (BLIP) and Z-sensitive Ionization- and Phonon-mediated
(ZIP) detectors
\cite{tomprl1,tomprl2,blipltd7,irwin2,alexsheffield,clarkepreprint}.
For both types, the drift field for the ionization measurement is
supplied by radially-segmented electrodes on the faces of
the disk-shaped crystals~\cite{tomltd8}.  In BLIP detectors, phonon
production is determined from the detector's calorimetric temperature
change.  In ZIP detectors, athermal phonons are collected to determine
phonon production and xy-position.  Detector performance is discussed
in detail elsewhere~\cite{blipltd7,alexsheffield,clarkepreprint}.

Photons cause most bulk electron recoils, while low-energy electrons
incident on the detector surfaces cause low-$Y$ electron recoils in a
thin surface layer (``surface events'').  Neutron, photon, and
electron sources are used to determine efficiencies for discrimination
between nuclear recoils and bulk or surface electron recoils.  Above
10 keV, CDMS detectors reject bulk electron recoils with $>$~99\%
efficiency and surface events with $>$~95\% efficiency.  CDMS
detectors that sense athermal phonons provide further surface-event
rejection based on the differing phonon pulse shapes of bulk and
surface events~\cite{clarkepreprint}.  This phonon-based surface-event
rejection alone is $>$~99.7\% efficient above 20 keV.

The 1-cm-thick, 7-cm-diameter detectors are stacked 3 mm apart
with no
intervening material.  This close packing
enables the annular outer ionization electrodes to shield the
disk-shaped inner electrodes from low-energy electron sources on
surrounding surfaces.  The probability that a surface event will
multiply scatter is also increased.

The low rate of WIMP interactions necessitates operation at a site
with low background-particle flux.  CDMS detectors are operated
beneath 16 meters-water-equivalent overburden, which stops
the hadronic component of cosmic-ray air showers and reduces the
muonic component by a factor of 5.  A custom, radiopure extension to a
modified Oxford S-400 dilution refrigerator provides a low-background
20 mK volume~\cite{icebox}.

Several layers of shielding surround the cryostat.  Outermost is a
$>$~99.9\% efficient plastic-scintillator veto to detect muons and thus
allow rejection of muon-coincident particles.  Inside the veto, a
15-cm-thick lead shield reduces the background photon flux by a factor
of 1000.  A 1-cm-thick shield made of ancient lead provides additional
photon shielding inside the cryostat~\cite{dasilvapb210}.  Samples of
all construction materials were screened to ensure low
radioactive contamination.  The measured event rate below 100 keV due
to photons is roughly 60~keV$^{-1}$~kg$^{-1}$~d$^{-1}$ overall and
2~keV$^{-1}$~kg$^{-1}$~d$^{-1}$ anticoincident with veto.

Neutrons with energies capable of producing keV nuclear recoils are
produced by muons interacting inside and outside the veto
(``internal'' and ``external'' neutrons, respectively).  The dominant,
low-energy ($< 50$ MeV) component of these neutrons is moderated by a
25-cm thickness of polyethylene between the outer lead shield and
cryostat~\cite{dasilvaneutron}.  However, high-energy external
neutrons may punch through the moderator.  A simulation of these
neutrons assumes the production spectrum given in~\cite{extneutspec}
and propagates them through the shield to the detectors.  The accuracy
of the simulation's propagation of neutrons is confirmed by the
excellent agreement of the simulated and observed recoil-energy
spectra due to veto-coincident and calibration-source neutrons.  A
large fraction of the external neutrons are vetoed: $\sim$40\% due to
neutron-scintillator interactions and an unknown fraction due to
associated hadronic showers.  This unknown fraction, combined with a
factor of $\sim$4 uncertainty in the production rate, makes it
difficult to accurately predict the absolute flux of unvetoed external
neutrons.  However, normalization-independent predictions of the
simulation, such as relative rates of single scatters and multiple
scatters, relative rates in Si and Ge detectors, and the shapes of
nuclear-recoil spectra, are insensitive to reasonable changes in the
neutron spectrum.

Two data sets are used in this analysis: one consisting of 33 live
days taken with a 100 g Si ZIP detector between April and July, 1998,
and another taken later with Ge BLIP detectors.  The Si run yields a
1.6~kg~d exposure after cuts.  The total low-energy electron
surface-event rate is 60~kg$^{-1}$~d$^{-1}$ between 20 and 100 keV.
Four nuclear recoils are observed in the Si data set.  Based on a
separate electron calibration, the upper limit on the expected number
of unrejected surface events is 0.26 events (90\%~CL).  These nuclear
recoils also cannot be due to WIMPs.  Whether their interactions with
target nuclei are dominated by spin-independent or spin-dependent
couplings, WIMPs yielding the observed Si nuclear-recoil rate would
cause an unacceptably high number of nuclear recoils in the Ge data
set discussed below.  Therefore, the Si data set, whose analysis is
described elsewhere~\cite{alexprl}, measures the unvetoed neutron
background.

Between November, 1998, and September, 1999, 96~live~days of data were
obtained using 3 of 4 165~g Ge BLIP detectors.  One detector is
discarded because it displays a high rate of veto-anticoincident
low-energy electron surface events, 230~kg$^{-1}$~d$^{-1}$ as compared
to 50~kg$^{-1}$~d$^{-1}$ for the other detectors (10 to 100~keV).
This detector suffered additional processing steps that may have
contaminated its surface and damaged its electrodes.  Data-quality,
nuclear-recoil acceptance, and veto-anticoincidence cuts reduce the
exposure (mass $\times$ time) by 45\%.  To take advantage of
close packing, analysis is restricted to events fully contained in the
inner electrodes, reducing the exposure further by a factor of 2.5 to
yield a final Ge exposure of 10.6~kg~d~\cite{r19prd}.

\begin{figure}[h]
\begin{center}
\epsfig{figure=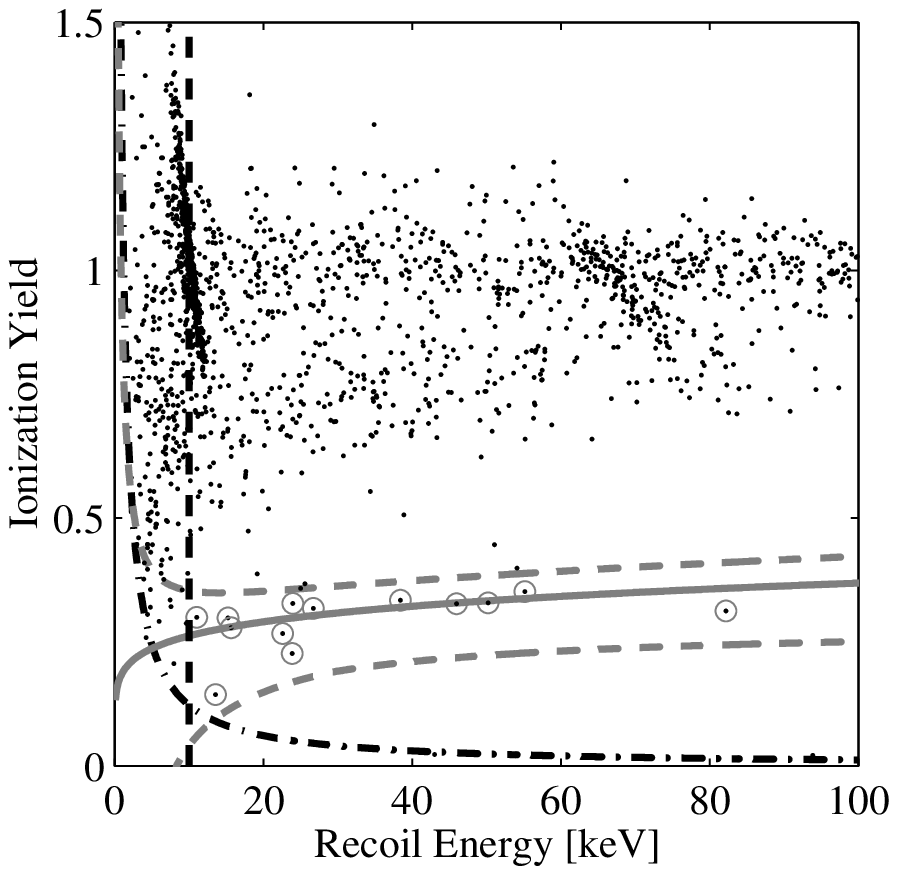, width=7.7cm}
\end{center}
\caption{Ionization yield ($Y$) vs. recoil energy for
veto-anticoincident single scatters contained in the inner electrodes
of the 3 uncontaminated Ge detectors.  Solid curve: expected position
of nuclear recoils.  Dashed curves: nominal 90\% nuclear-recoil
acceptance region.  Dashed line: 10~keV analysis threshold.
Dashed-dotted curve: threshold for separation of ionization signal
from amplifier noise.  Circled points: nuclear recoils.  The presence
of 3 events just above the acceptance region is compatible with 90\%
acceptance.}
\label{highbiasyplot}
\end{figure}

Figure \ref{highbiasyplot} shows a plot of ionization yield vs. recoil
energy for the Ge data set.  Bulk electron recoils
lie at ionization yield $Y \simeq 1$.  Low-energy electron events form a
distinct band at $Y\sim 0.75$, leaking into the nuclear-recoil
acceptance region below 10 keV.

\begin{figure}
\begin{center}
\epsfig{figure=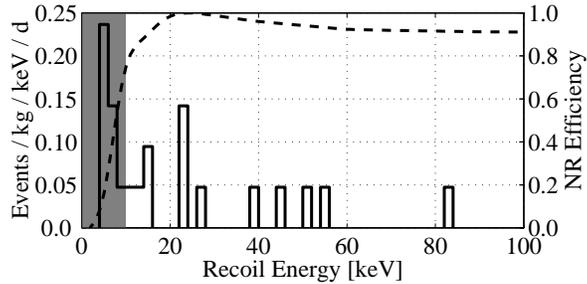, width=7.7cm}
\end{center}
\caption{Solid: histogram of nuclear recoils observed in the inner
electrodes of the 3 uncontaminated Ge detectors (left-hand
scale). Shaded: 10 keV analysis threshold.  Dashed: peak-normalized
nuclear-recoil efficiency (right-hand scale).}
\label{spec}
\end{figure}

Figure \ref{spec} displays the recoil-energy spectrum of unvetoed
nuclear recoils for the Ge
data set.  Only single scatters (events triggering a single detector)
are shown; the WIMP multiple-scatter rate is negligible.  An analysis
threshold of 10 keV, well above trigger thresholds, is imposed.  This
choice reduces the data set's sensitivity but simplifies analysis by
rendering low-energy electron misidentification negligible.  The
nuclear-recoil efficiency is determined using calibration-source
neutrons and its stability is monitored using veto-coincident
neutrons.  Thirteen unvetoed nuclear recoils are observed in the
10.6~kg~d exposure between 10 and 100 keV; this rate is similar to
that expected for the WIMP signal claimed by the DAMA experiment
\cite{DAMA2000,DAMA1998}.  However, much evidence indicates that the
CDMS nuclear recoils are caused by neutrons rather than WIMPs.

\begin{figure}
\begin{center}
\epsfig{figure=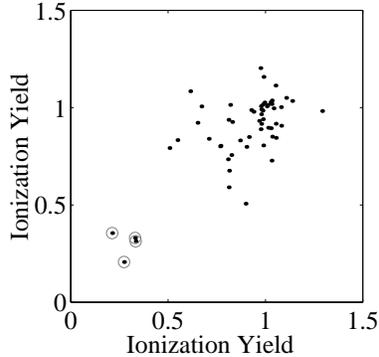, width=5cm}
\end{center}
\caption{Scatter plot of ionization yields for multiple scatters in
the 3 uncontaminated Ge detectors with at least 1~inner-electrode
scatter and both scatters between 10 and 100~keV.  Circled events are
tagged as nuclear recoils in both detectors.  Bulk recoils and surface
events lie at $Y \simeq 1$ and $Y \sim 0.75$, respectively.}
\label{multiy2}
\end{figure}

Figure \ref{multiy2} displays a scatter plot of ionization yields for
multiple scatters.  The observation of 4 Ge multiple-scatter nuclear
recoils is the primary evidence for the neutron interpretation.  It is
highly unlikely that these events are misidentified low-energy
electron events.  Figures \ref{highbiasyplot} and \ref{multiy2}
demonstrate excellent separation of low-energy electron events from
nuclear recoils.  Analysis using events due to 
electrons emitted by the contaminated detector yields an upper
limit of 0.05 misidentified multiple-scatter low-energy electron
events (90\%~CL).

All other pieces of evidence are also consistent with the neutron
interpretation.  First, the 4 nuclear recoils observed in the Si data
set cannot be interpreted as WIMPs or surface events.  Second, there
is reasonable agreement between predictions from the Monte Carlo
simulation and the relative observed numbers of 4 Ge multiple
scatters, 4 Si single scatters, and 13 Ge single scatters.
Normalizing the simulation by the 17 total Ge nuclear-recoil events yields
2.7 expected Si single scatters and 1.3 expected Ge multiple scatters.
Statistically, the expected neutron background should result in a less
likely combination of Ge single scatters, Ge multiple scatters, and Si
single scatters 23\% of the time.  Finally, a Kolmogorov-Smirnov test
indicates that the deviation between the observed and simulated
nuclear-recoil spectral shapes would be larger in 28\% of experiments.

\begin{figure}
\begin{center}
\epsfig{figure=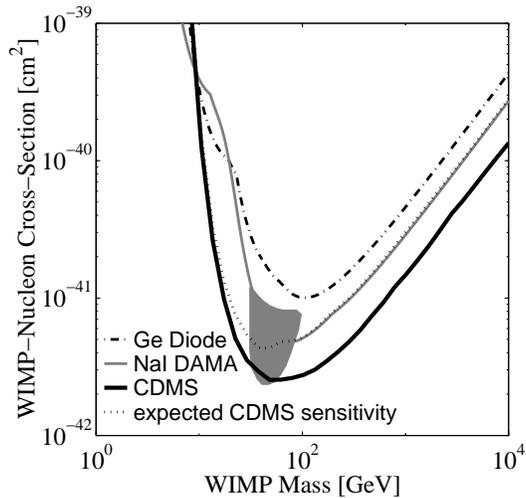, width=7cm}
\end{center}
\caption{Spin-independent $\sigma$ vs. $M$.  The regions above the
curves are excluded at 90\%~CL.  Solid dark curve: limit from this
analysis.  Dotted curve: CDMS expected sensitivity (median simulated
limit) given the observed neutron background.  Because the number of
multiple scatters observed is larger than expected, the limit from
this analysis is lower than the median simulated limit.  Solid light
curve: DAMA limit using pulse-shape analysis~\protect\cite{DAMApsa}.
Shaded region: DAMA 3$\sigma$ allowed region as described in
text~\protect\cite{DAMA2000}.  Dashed-dotted curve: Ge diode limit,
dominated by~\protect\cite{hm}.  All curves are normalized
following~\protect\cite{lewin}, using the Helm spin-independent
form-factor, $A^2$~scaling, WIMP characteristic velocity $v_0 = 220$
km s$^{-1}$, mean Earth velocity $v_E = 232$ km s$^{-1}$, and $\rho =
0.3$~GeV~c$^{-2}$~cm$^{-3}$.}
\label{limitplot}
\end{figure}

The 90\%~CL excluded region for the WIMP mass $M$ and the
spin-independent WIMP-nucleon elastic-scattering cross-section
$\sigma$ is derived using an extension of the approach of Feldman and
Cousins~\cite{feldmancousins}.  The above arguments require accounting
for the component of the observed Ge single scatters (with energies
$E_{\mathrm i}$) that is due to the unvetoed neutron flux $n$.  This
flux is constrained by the number $N_{\mathrm m}$ of multiple scatters
in Ge and the number $N_{\mathrm Si}$ of nuclear recoils in Si.  To
determine the 90\%~CL excluded region in the plane of $M$ and $\sigma$
alone, the parameter $n$ is projected out.  For a grid of physically
allowed values of $M$, $\sigma$, and $n$, the expected distribution of
the likelihood ratio $R = {\cal L}( E_{i}, N_{\mathrm m}, N_{\mathrm
Si} | \,\sigma, M, \tilde{n})/{\cal \widehat{L}}$ is calculated by
Monte Carlo simulation in order to determine the critical parameter
$R_{90}$ such that 90\% of the simulated experiments have $R>R_{90}$.
Here $\tilde{n}$ is the value of $n$ that maximizes the likelihood
${\cal L}$ for the given parameters $M$ and $\sigma$ and the
observations.  ${\cal \widehat{L}}$ is the maximum of the likelihood
for any physically allowed set of parameters.  The WIMP-nucleon
cross-section $\sigma$ is converted to a WIMP-nucleus cross-section
assuming $A^2$~scaling with target nuclear mass.
This scaling is valid for models of supersymmetric WIMPs
currently favored~\cite{jkg,ellis,gondolo,bottino}.
The 90\%~CL region excluded by the observed data set, with ratio
$R_{\mathrm data}$, consists of all parameter space for which
$R_{\mathrm data}\leq R_{90}$.  Figure \ref{limitplot} displays the
lower envelope of points excluded for all values of $n$.
Because all the nuclear recoils may be neutron scatters, 
$\sigma = 0$ is not excluded.

This limit excludes new parameter space for WIMPs with $M>$~10 GeV
c$^{-2}$, some of which is allowed by supersymmetry~\cite{gondolo}.
The data are compatible with the DAMA/NaI-0 exclusion limit based on
pulse-shape analysis~\cite{DAMApsa}.  However, these data exclude, at
$> 84\%$~CL, the entire region allowed at 3$\sigma$ by the DAMA/NaI-1 to 4
annual modulation signal~\cite{DAMA2000}.  This region, given by the
$v_0 = 220$~km~s$^{-1}$ curve in Figure 4a of Ref.~\cite{DAMA2000}, is
used because it is determined solely from the annual modulation
signal.  The data presented here also exclude the analogous 2$\sigma$
allowed region for DAMA/NaI-1 to 2 at $> 96$\%~CL~\cite{DAMA1998}.
Although without theoretical support, non-$A^2$ scaling may allow the
two results to be compatible.

We thank Paul Luke of LBNL for his advice regarding surface-event
rejection.  We thank the engineering and technical staffs at our
respective institutions for invaluable support.  This work is
supported by the Center for Particle Astrophysics, an NSF Science and
Technology Center operated by the University of California, Berkeley,
under Cooperative Agreement No. AST-91-20005, by the National Science
Foundation under Grant No. PHY-9722414, by the Department of Energy
under contracts DE-AC03-76SF00098, DE-FG03-90ER40569,
DE-FG03-91ER40618, and by Fermilab, operated by the Universities
Research Association, Inc., under Contract No. DE-AC02-76CH03000 with
the Department of Energy.


\bibliographystyle{prsty}

\newpage

\end{document}